\documentclass[amsmath, amssymb,10pt, aps, prb, twocolumn, notitlepage, longbibliography, superscriptaddress]{revtex4-1}

\usepackage{graphicx}
\usepackage{calc}
\usepackage{braket}
\usepackage{slashed}
\usepackage{wasysym}
\usepackage{dsfont}
\usepackage{amsthm,amsmath,amsfonts,amssymb,verbatim,color}
\usepackage{lipsum}
\usepackage{bm}
\usepackage{epsfig,slashed}
\usepackage{hyperref}
\usepackage{flafter}
\usepackage{booktabs}
\usepackage{makecell} 
\usepackage{bbm} 
\usepackage[export]{adjustbox} 

\newcommand{\bk}{{\bf k}}
\newcommand{\bP}{{\bf P}}
\newcommand{\bp}{{\bf p}}

\newcommand{\br}{{\bf r}}

\def\brho{{\boldsymbol \rho}}

\newcommand{\hG}{{\hat G}}

\newcommand{\hh}{{\hat h}}

\newcommand{\hn}{{\hat n}}

\newcommand{\hPsi}{{\hat\Psi}}
\newcommand{\hpsi}{{\hat\psi}}
\newcommand{\hsigma}{{\hat\sigma}}
\newcommand{\hxi}{{\hat\xi}}

\newcommand{\cD}{{\cal D}}

\newcommand{\cH}{\hat{\cal H}}

\def\holOne{\mathds{1}}

\newcommand{\dis}{\text{dis}}

\begin{document}
\title{Interactions-disorder duality and critical phenomena in nodal semimetals, dilute gases and other systems}

\author{Shijun Sun}
\affiliation{Physics Department, University of California, Santa Cruz, California 95064, USA}

\author{Sergey Syzranov}
\affiliation{Physics Department, University of California, Santa Cruz, California 95064, USA}

\begin{abstract}
We investigate classes of interacting systems that allow for a mapping to disordered non-interacting systems.
As we show, such a mapping is possible for interacting systems with a suppressed density of states at the chemical potential,
leading to suppressed screening, and systems near BCS-type instabilities.
The mapping can also be applied qualitatively to other classes of systems that are not exactly dual to each other.
The established duality suggests a new approach to analytical and numerical studies of many-body 
and disorder-driven phenomena in a variety of systems and allows to predict, e.g., new phase transitions dual to the previously
known ones.
Using the established duality, we predict new disorder-driven transitions in nodal-line semimetals and systems
with long-range hopping dual to, respectively, the BCS and BEC-vacuum transitions in interacting systems
and new interaction-driven transitions dual to previously known non-Anderson disorder-driven transitions.
The established principle can also be used to classify and describe phase transitions in dissipative systems described by non-Hermitian Hamiltonians.
\end{abstract}


\maketitle


Describing many-body interacting systems is one of the greatest challenges in physics.
Often, existing analytical approaches are insufficient to accurately describe many-body effects, such as high-temperature superconductivity, interaction-driven metal-insulator transition and magnetic instabilities.

Simulating such systems numerically is also a formidable task, especially in the case of fermionic (quasi-)particles, which display the notorious 
sign problem~\cite{BlackenbeclerScalapino:firstSignProblem,Hirsch:firstSignProblem,WhiteScalapinoScalettar:signProblem} leading to a rapid growth of the
computation time with the number of particles.
By contrast, single-particle problems, even in the presence of quenched disorder, may be comparatively 
easily simulated, e.g., by diagonalising the Hamiltonian of the system.

Disorder-averaged observables in single-particle models, however, are described by interacting field theories, in the supersymmetric~\cite{Efetov:book}, Keldysh~\cite{Kamenev:book} or replica~\cite{BelitzKirkpatrick:review} representations.
Such theories have the same form as the field theories of interacting disorder-free systems but have additional structures in, respectively, boson-fermion, Keldysh or replica subspaces.
It is natural to expect, therefore, that a certain class of many-body systems may be mapped to single-particle disordered models.

Such a mapping would allow one for easier numerical simulations of many-body phenomena,
reducing them to single-particle problems.
Furthermore, it would allow to predict new many-body (disorder-driven) phenomena,
e.g. phase transitions, 
dual to previously known disorder-driven (many-body) effects.
The purpose of this paper is to explore such a duality between interacting and disordered non-interacting
systems and its consequences for critical phenomena.

\begin{figure}[ht!]
	\centering
	\includegraphics[width=0.35\linewidth]{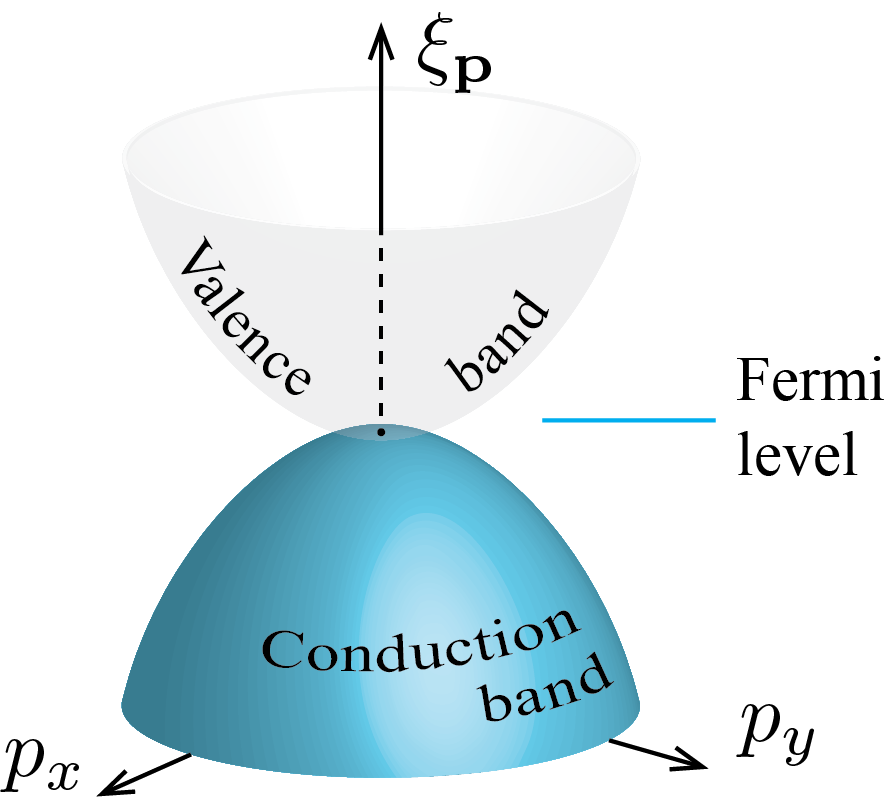}
	\caption{\label{fig:semimetalbands}
		The band structure of a nodal semimetal. For the power-law dispersion $\propto p^\alpha$, the density of states 
		$\rho(\varepsilon)\propto |\varepsilon|^{\frac{d}{\alpha}-1}$ is suppressed in high dimensions $d>\alpha$, which leads to 
	a suppressed screening of the interactions.}
\end{figure}

We demonstrate that such a duality indeed exists for broad classes of interacting and disordered systems and interaction-
and disorder-driven phenomena.
Such phenomena include, but are not limited to, ultraviolet (UV) effects~\cite{Syzranov:review,ZhuSyzranov:BCSnodalSemimetal},
i.e. transport effects and phase transitions that come from 
quasiparticle scattering through states far from the chemical potential (in a momentum band exceeding the inverse mean free path $\ell^{-1}$).
Such phenomena include I) effects in nodal semimetals and systems in
sufficiently high dimensions (unconventional quantum corrections to transport~\cite{AleinerEfetov:graphene}, non-Anderson disorder-driven
transitions~\cite{Syzranov:review}, unconventional superconductive instabilities~\cite{Kozii:superconductivityDirac,MaciejkoNandkishore:WeylWrongRG}, etc.) and II) BCS-type interaction-driven instabilities with 
a single interaction channel. The duality transformation that we derive 
can also be applied to other systems to reveal new phenomena that are
not exactly dual but similar to the previously known ones.

The duality transformation we develop in this paper maps a $d$-dimensional interacting disorder-free system to 
a $d+1$-dimensional non-interacting systems with quenched disorder and an additional (pseudospin) degree of freedom equivalent to a spin-$1/2$. The mapping is exact if the screening of the interactions can be neglected,
as is the case for systems I and II mentioned above.

%
%

The existence of such a duality mapping for nodal semimetals
is consistent with the well-known~\cite{Hirsch:firstSignProblem} absence of the numerical sign problem in systems with particle-hole symmetry. 
Indeed, nodal semimetals with symmetric valence and conduction bands and chemical potentials close to the band-touching point (see Fig.~\ref{fig:semimetalbands}) have (approximate) particle-hole symmetry.

The mapping we derive suggests, in particular, that interacting electronic systems with a vanishing DoS at the Fermi level exhibit
interaction-driven phase transitions in the universality classes of the non-Anderson disorder-driven phase transitions~\cite{Fradkin1,Fradkin2,Syzranov:review,Rodriguez:firstPowerLaw, Malyshev:firstPowerLaw,Syzranov:WeylTransition,Syzranov:unconv}
that have been established, at the perturbative level,
to take place in non-interacting semimetals in high spacial dimensions, exemplified by 3D Weyl semimetals.
Using the duality demonstrated in this paper, we find a new non-Anderson disorder-driven phase transition
dual to the previously known ``vacuum-BEC'' transition~\cite{Uzunov:BCSRG,NikolicSachdev:RGFeshbah,GurarieRadzihovsky:FeshbahReview}.
Furthermore, we predict new disorder-driven transitions dual to the interaction-driven 
BCS-type transitions (e.g. superconductive and excitonic instabilities).
Such transitions lead, for example, to the critical scaling of the density of states 
in disordered nodal-line semimetals~\cite{ZhuSyzranov:BCSnodalSemimetal}. The duality can be extended further to the 
case of non-Hermitian Hamiltonians and phase transitions.

{\it Summary of the mapping.}
We consider $d$-dimensional interacting systems described by Hamiltonians of the form
\begin{align}
	\cH= &\int \hPsi^\dagger(\br) \xi_{\hat \bp} \hPsi (\br)d^d\br
	\nonumber\\
	&-\frac{1}{2}\int \hPsi^\dagger(\br)\hPsi^\dagger(\br^\prime)U(\br-\br^\prime)\hPsi(\br^\prime)\hPsi(\br)d^d\br d^d\br^\prime,
	\label{HamiltonianInteracting}
\end{align}
where $\hPsi^{(\dagger)}$ are the particle operators; $\xi_{\hat \bp}$ is the single-particle dispersion;
$\hat \bp=-i\partial_\br$ is the momentum operator (hereinafter $\hbar=1$) and the potential $U(\br-\br^\prime)$
describes the interactions.
The particles may be bosonic or fermionic.
While we consider, for simplicity, spinless particles, the mapping described in this paper may be generalised to account for an arbitrary spin structure of the dispersion and the interaction.


We assume that the screening is either suppressed or has no qualitative effects on the observables and phenomena 
of interest, i.e. does not change the universality class of a phase transition.
This is often the case for phenomena dominated by the {\it ultraviolet} energy and momentum scales.


For example, in 
nodal semimetals and gases with the power-law dispersion $\xi_\bp \propto p^\alpha$ (possibly with an additional structure in the spin/valley space) in high dimensions $d>\alpha$,
the screening will be suppressed due to the suppressed density of states, $\rho(\varepsilon)\propto 
|\varepsilon|^{\frac{d}{\alpha}-1}$, at the node or a band touching point ($k=0$).
$\alpha=1$ corresponds to graphene in the dimension $d=2$ and to 3D Weyl/Dirac semimetals in the dimension $d=3$. The case $\alpha=2$ describes
semiconductors and parabolic semimetals. Dispersions with a continuously tunable parameter $\alpha$ may be realised in systems of trapped ultracold particles~\cite{Monroe:longrange,Islam:longrange,Blatt:chain1,Blatt:chain2} and
superconductive films~\cite{TikhonovFeigleman:PowerLawSupercond}. Furthermore, $d$-dimensional systems with
long-range hopping $\propto1/r^{d-\alpha}$, also realised with trapped ultracold particles, 
are dual to systems with the power-law dispersion $\xi_\bp \propto p^\alpha$~\cite{SyzranovGurarie:duality}.

For some phenomena, such as BCS-type instabilities (superconductive or exciton-condensation instabilities)
or leading-order correlators of electron densities,
screening is not important even in the presence of a large Fermi surface.

For the case of {\it attractive interactions}, on which we focus in most of this paper, 
the behaviour of observables in an interacting disorder-free $d$-dimensional system described by the  Hamiltonian~\eqref{HamiltonianInteracting} 
can be mapped to the behaviour of disorder-averaged observables 
in a $d+1$-dimensional non-interacting semimetal with the Hamiltonian
\begin{align}
	\hh=\hsigma_z\xi_{\hat \bp}+\hsigma_y p_{d+1}+\hsigma_z u(\brho),
	\label{HamiltonianDisordered}
\end{align}
where $\hsigma_x$, $\hsigma_y$ and $\hsigma_z$ are the Pauli matrices corresponding to a degree of freedom equivalent to a spin-$1/2$,
hereinafter referred to as {\it pseudospin}; 
$\bp$ is the momentum of the particle along the first $d$ dimensions; $p_{d+1}$ is the component of momentum along the $d+1$-st dimension; $u(\brho)$ is a random potential in the $d+1$-dimensional space whose correlator $-D(\br,\tau;\br^\prime,\tau^\prime)$ is given by the propagator of the interactions in Eq.~\eqref{HamiltonianInteracting}. 
For short-ranged interactions, the strength of both the interactions and the random potential are
described by one coupling constant
\begin{align}
	g=\int U(\br-\br^\prime) d^d\br^\prime
	= \int \left<u(\brho)u(\brho^\prime)\right>_{\text{dis}}d^{d+1}\brho^\prime.
	\label{Coupling}
\end{align}
Along the $d+1$-st dimension, the size of the system described by the Hamiltonian~\eqref{HamiltonianDisordered}
is given by $\ell_{d+1}=1/T$,
where $T$ is the temperature of the interacting system with the Hamiltonian~\eqref{HamiltonianInteracting},
and (anti-)periodic conditions are imposed for (fermionic) bosonic particles.

The mapping can be similarly carried out for {\it repulsive interactions}, with the Hamiltonian~\eqref{HamiltonianDisordered} replaced by
\begin{align}
	\hh_{\text{repulsive}}=\hsigma_z\xi_{\hat \bp}+\hsigma_y p_{d+1}+\hsigma_y u(\brho).
	\label{HamiltonianDisordered2}
\end{align}

Each observable in the interacting model~\eqref{HamiltonianInteracting} corresponds to a disorder-averaged quantity 
in the disordered non-interacting model~\eqref{HamiltonianDisordered}.
For example, the average density $\hn(\br)=\hPsi^\dagger(\br)\hPsi(\br)$ of the interacting particles
matches, as a function of the disorder/interaction strength (e.g., coupling $g$), the disorder-averaged quantity 
\begin{align}
	\rho_s(\brho)=\frac{1}{4}\mathrm{Tr}\left[\hsigma_z\hG^R(\brho,\brho,0)+\hsigma_z\hG^A(\brho,\brho,0)\right]
	\label{RhoDefinition}
\end{align}
in the dual disordered non-interacting system, where $\mathrm{Tr}$ is taken over the pseudospin degree of freedom
and $\hG^R(\brho,\brho,E)$ and $\hG^A(\brho,\brho,E)$ are the matrices of the retarded and advanced Green's functions
of the particles in the pseudospin space. Similar correspondence can be established for other observables,
such as currents and spin densities.

\begin{figure}[ht!]
	\centering
	\includegraphics[width=0.9\linewidth]{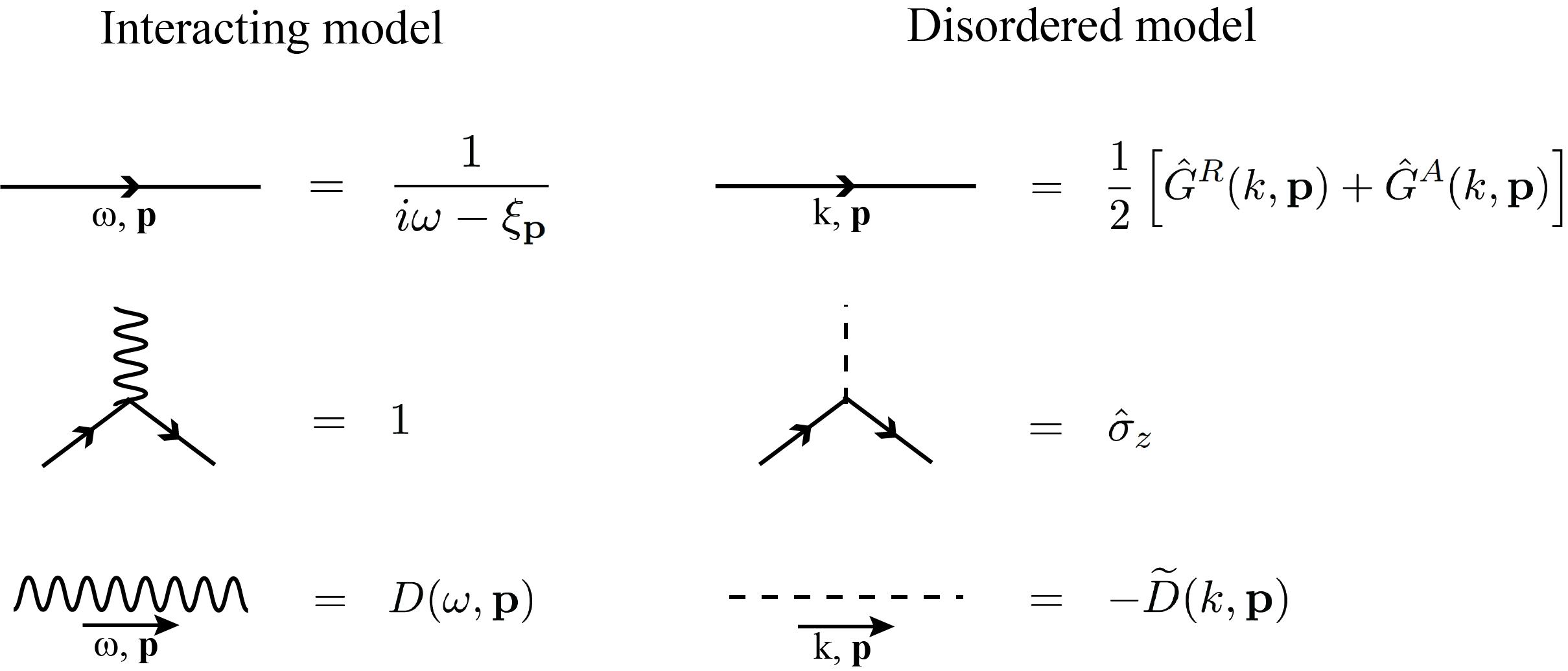}
	\caption{\label{fig:elementsDiagrammatics} Elements of the diagrammatic technique for the interacting disorder-free (left) and non-interacting 
		disordered (right) models in momentum space that illustrate perturbative equivalence between the two classes of systems
	}	
\end{figure}


{\it Types of dual systems.} 
There are multiple types of disordered systems described by
the Hamiltonian~\eqref{HamiltonianDisordered} [Eq.~\eqref{HamiltonianDisordered2}] and dual interacting systems.
They include several broad classes:
I) if $\xi_\bp$ vanishes at a point ($\bp=0$), the duality provides a mapping between an interacting nodal-point
semimetal and a disordered anisotropic nodal semimetal in a higher dimension;
II) a generic 2D interacting metal is mapped to a 3D disordered nodal-line semimetal;
III) at very high temperatures $T$ in the interacting system,
the motion of the dual disordered system 
is strongly constrained (quantised) along the $d+1$-st dimension and is reduced to the $d$-dimensional motion
of a particle in a random potential with the generic Hamiltonian $\hh_\text{eff}=\xi_\bp+u(\brho)$.

For all of these types of systems, the derived mapping
reveals new critical phenomena dual to previously known ones, as detailed below. 
For systems in groups I and II, the duality is exact under the conditions discussed in the introduction.
Below, we review some of the phenomena revealed for such systems by the duality mapping derived in this manuscript.  
For systems in group III, the assumption about the negligibility of interactions is, in general, not fulfilled.
Nevertheless, the mapping allows us to predict a high-temperature
interaction-driven transition in a system with the 
power-law dispersion $\xi_\bp\propto p^\alpha$ associated 
by the duality mapping with the non-Anderson disorder-driven transitions~\cite{Syzranov:review} for particles with the same dispersion in the same dimension.
We will provide a detailed description of such interaction-driven transitions elsewhere~\cite{SuSyzranov:HighT}.


{\it Perturbative derivation of duality.} 
The summarised duality
can be rigorously verified to all orders of the perturbation theory
under the made approximations, with the corresponding elements of the diagrammetic technique shown in Fig.~\ref{fig:elementsDiagrammatics}.

\begin{figure}[hb!]
	\centering
	\includegraphics[width=\linewidth]{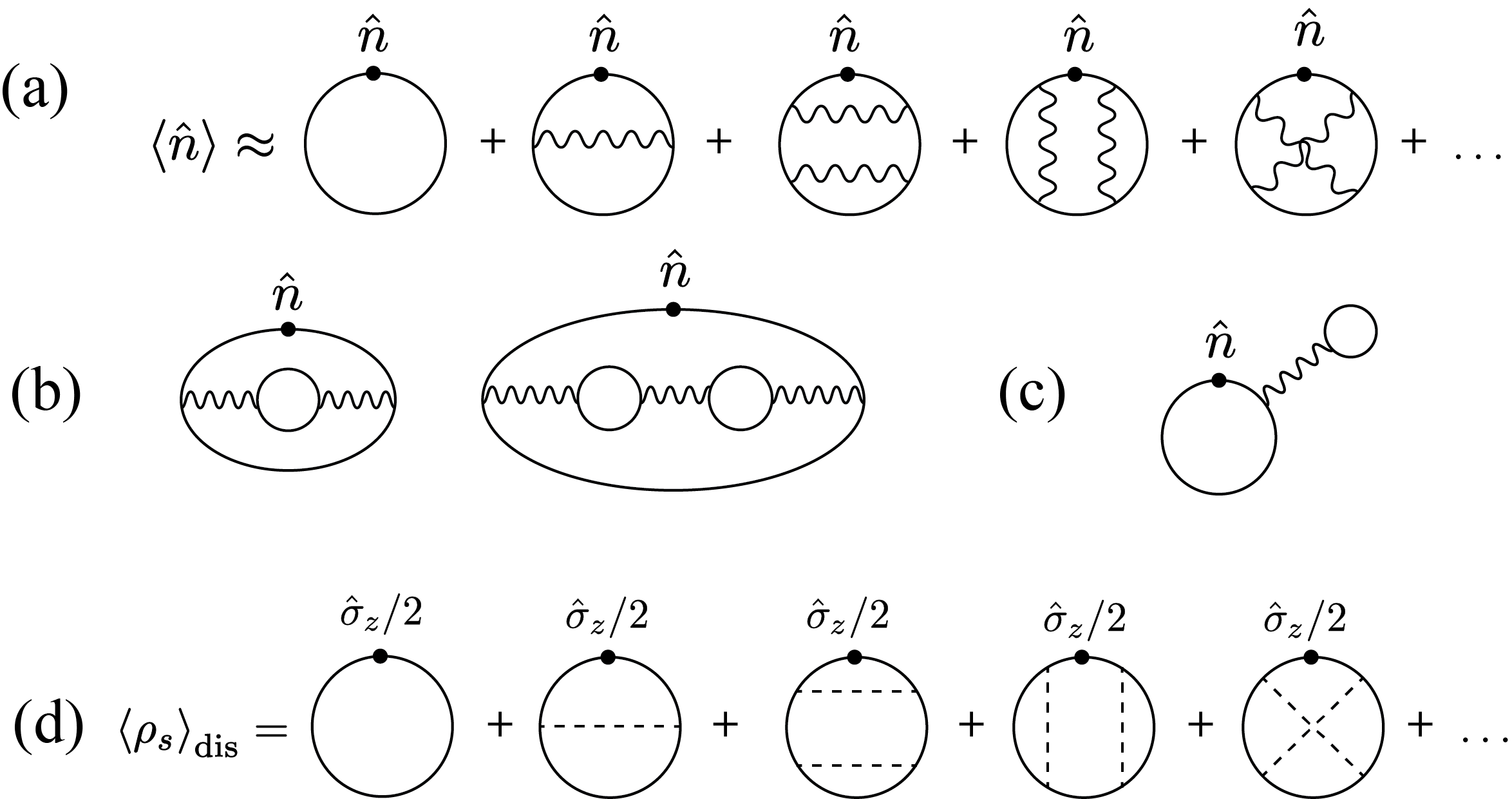}
	\caption{
		\label{fig:maindiagrams}
		Diagrams for observables in dual interacting disorder-free and non-interacting disordered systems.
		(a) Contributions to the concentration of interacting particles. (b)-(c) Examples of neglected contributions;
		diagrams (b) are neglected due to the screening suppression; (c) is the Hartree contribution.
		(d) Corresponding contributions to the disorder averaged quantity $\rho_{s}$, given by Eq.~\eqref{RhoDefinition}
		in the disordered non-interacting system. 
	}
\end{figure}

For example, perturbative contributions to the average density of particles $\hn$ in 
the interacting system described by Hamiltonian~\eqref{HamiltonianInteracting} are shown in Figs.~\ref{fig:maindiagrams}a-c.
The diagrams in Fig.~\ref{fig:maindiagrams}a include, apart from the interaction propagators (wiggly lines), only one loop of particle propagators (solid lines).
Contributions with additional loops of propagators,
exemplified by diagrams~\ref{fig:maindiagrams}b-c, can be neglected.
Some of those neglected contributions (Fig.~\ref{fig:maindiagrams}b) describe the screening of the interactions.
The others contain a loop connected to the rest of the diagram by a single interaction propagator
(Hartree-type contributions, shown in Fig.~\ref{fig:maindiagrams}c) which may be absorbed in the definition of the chemical
potential and has no qualitative effects. 

The remaining diagrams in the interacting systems [Fig.~\ref{fig:maindiagrams}a] correspond to
topologically equivalent diagrams in the dual disordered systems [Fig.~\ref{fig:maindiagrams}c].
In Supplemental Material~\cite{SupplementalInformation}, we provide a detailed comparison of the respective 
diagrams and demonstrate their equal values to all orders of the perturbation theory.


{\it Heuristic argument.}
The mapping can be understood intuitively from the following heuristic argument.
The partition function of a $d$-dimensional system
with the attractive-interaction term decoupled by 
the Hubbard-Stratonovich field $\phi(\br,\tau)$
\begin{align}
	Z=\int \cD\bar{\psi}\cD\psi\cD\phi \exp\left[-\int\bar{\psi}\left(\partial_\tau+\xi_\bp+\phi\right)\psi\, d^d\br d\tau
	\right.
	\nonumber\\
	\left.
	-\frac{1}{2}\int \phi(\tau,\br)U^{-1}(\br,\br^\prime)\phi(\tau,\br^\prime)d\tau d^d\br d^d\br^\prime\right]
	\label{TheoryDecoupled}
\end{align}
resembles the partition function of a disordered $d+1$-dimensional system, where the Matsubara time $\tau$ 
is considered as an extra coordinate, with a non-Hermitian Hamiltonian $\tilde{h}=\partial_\tau+\xi_\bp+\phi$ and $\phi$ playing the role of the
disorder potential.

The procedure of ``Hermitisation''~\cite{Efetov:Hermitisation,LuOhtsukiShindou:nonHermAnderson,Feinberg:Hermitisation,Gong:nonHermitianReview,Kawabata:nonHermitianSymmetries,Bergholtz:nonHermitianReview}
can be applied then to associate the non-Hermitian operator $\tilde{h}=\partial_\tau+\xi_\bk+\phi$
with its ``Hermitised'' version ${\hat{h}}_{\text{herm}}=\hsigma_x p_\tau+\hsigma_y \xi_\bk+\hsigma_y \phi$
 in the doubled Hilbert space,
where $p_\tau=-i\partial_{\tau}$. The ``Hermitised'' Hamiltonian matches, up to a rotation of the pseudospin
basis, the Hamiltonian~\eqref{HamiltonianDisordered}. The properties of this Hamiltonian
are similar to those of $\tilde{h}$ and allow for obtaining related observables.

Another distinction of the action~\eqref{TheoryDecoupled} from that of a disordered system is 
that it uses Grassmann (real) fields $\bar{\psi}$ and $\psi$ for fermionic (bosonic) particles
not including the fermion-boson, replica or Keldysh subspaces inherently present
in field theories of disordered systems. As a result, it allows, in general, for ``loop'' contributions to observables 
that account for, e.g., the screening of the interactions and that are absent in disordered field theories~\cite{AGD,Efetov:book,AltlandSimons:book}. Under the assumptions we make in this paper, however, such contributions are negligible.


{\it Duality between a 2D metal and 3D nodal-line semimetal.}
The dispersion $\xi_\bp$ is measured from the chemical potential and, in general,
vanishes on a finite surface, $\xi_\bp=0$, in momentum space. 
For a 2D interacting  metal, this surface is a line,
and the dual disordered Hamiltonian~\eqref{HamiltonianDisordered}
describes a 3D nodal-line semimetal, i.e. a semimetal
with two bands touching along a line in momentum space~\cite{Armitage:WeylReview}.   

Although the density of states is not suppressed for a 2D metal,
the duality can be applied if the screening of the interactions has no qualitative effects, e.g. does not change 
the universality class of a phase transition, such as at the superconductive (BCS-type) instability.
We predict, therefore, that a 3D nodal-line semimetal exhibits a disorder-driven transition dual to the BCS transition
in a 2D semimetal.
A microscopic derivation of such a transition is presented in a separate work~\cite{ZhuSyzranov:BCSnodalSemimetal}.


{\it Dilute gases and nodal-point semimetals.}
A broad class of interacting systems that satisfies the assumptions used to derive the duality
corresponds to the dispersion
$\xi_\bp$ that vanishes at a point ($\bp=0$) in momentum space, e.g., at the bottom of the dispersion in an interacting gas
or near the band touching point in a nodal-point semimetal (for example, Dirac, Weyl or parabolic semimetal).
For the power-law vanishing, 
$\xi_\bp\propto p^\alpha$, a $d$-dimensional interacting systems is mapped to an anisotropic disordered
nodal-point semimetal with the dispersion $\xi_\bp\hsigma_z+p_{d+1}\hsigma_y$.

Interacting nodal-point
semimetals display a variety of instabilities 
at low temperatures (see, for example, Refs.~\onlinecite{MengBalents:WeylSuperconductors,MaciejkoNandkishore:WeylInstabilities,Wei:excitonic,Wei:WSMOddInstability,ChoMoore:WSMSupercodInstability,TrescherKnolle:WSMTypeTwoInstabilities,Shi:CDWinTaSeI}), such as 
superconductive, magnetic and charge-density-wave phase transitions. By contrast, non-interacting disordered systems
are commonly believed to exhibit only one phase transition: the Anderson localisation-delocalisation transition.
However, it has been demonstrated, at the perturbative level, that semimetals and semiconductors with the power-law dispersion $\propto p^\delta$ in 
high dimensions $\tilde{d}>2\delta$
exhibit additional disorder-driven transitions (see Ref.~\onlinecite{Syzranov:review} for a review) in non-Anderson universality classes~\footnote{Exponentially rare non-perturbative effects (rare-region effects) may convert such transitions
	 to sharp crossovers, as discussed recently in the context of 3D Weyl semimetals~\cite{Wegner:DoS,Suslov:rare,Nandkishore:rare}\cite{Syzranov:unconv,PixleyHuse:missedPoint,WilsonPixley:rareSingleCone,PixleyWilson:reviewRare}. However, dual interacting transitions are true phase transitions as the described
	  interactions-disorder duality applies only at the perturbative level and does not extend to rare-region effects}. 
These non-Anderson disorder-driven transitions may have diverse properties depending on the symmetries of disorder and of the band structure, 
and, under some approximations~\cite{Note1}, display a critical behaviour of the density of states~\cite{ShindouMurakami,GoswamiChakravarty,RyuNomura,KobayashiOhtsukiHerbut:scaling,Syzranov:WeylTransition,Syzranov:unconv,LouvetFedorenko:theirFirst,RoyJuricic:superuniversality,Syzranov:TwoLoop,PixleyHuse:missedPoint,Sbierski:superAccurate,BeraRoy:inaccurateNumerics,Sbierski:WeylVector,LiuOhtsuki:LateNumerics,Balog2018:porousMedium,LuoOhtsukiShindou:multicriticality}, in contrast with the Anderson transitions.


The duality allows us to
predict a new non-Anderson disorder-driven transition, distinct from all the previously studied
transitions, in semimetals with the Hamiltonian~\eqref{HamiltonianDisordered}
in which the dispersion $\xi_\bp$ vanishes at small momenta. This transition is dual to the so-called
vacuum-BEC transition~\cite{Uzunov:BCSRG} 
in systems of interacting bosons with attractive interactions~\cite{GurarieRadzihovsky:FeshbahReview,NikolicSachdev:RGFeshbah,VailetteRadzihovsky:largeN}.

For $\xi_\bp \propto p^\alpha$ and short-range interactions (disorder), 
the instability for both interacting and disordered systems can be demonstrated by
the renormalisation-group (RG) analysis of the dimensionless coupling constant
\begin{align}
	\gamma=\frac{\zeta S_d}{2(2\pi)^d}gK^{d-\alpha},
\end{align}
where $S_d$ is the volume of a unit $d$-dimensional sphere; $K$ is the ultraviolet momentum cutoff, e.g. the characteristic size of the band in momentum space; $\zeta$ is a factor of order unity that depends on the spin and valley structure of the dispersion near the node or the band edge ($\zeta=1$ for $\xi_\bk=k^\alpha$).
Upon integrating out the highest momentum modes of the particles in both systems~\cite{SupplementalInformation}, the flow
of the dimensionless coupling is given by the (exact~\cite{SupplementalInformation}) RG equation
\begin{align}
	\partial_l \gamma = (\alpha-d)\gamma + \gamma^2,
	\label{RGflowCoupling}
\end{align}
which signals a phase transition (in systems with attractive interactions, $g>0$) in high dimensions $d>\alpha$ at the critical coupling $\gamma_c=d-\alpha$.

For interacting bosons, the corresponding transition occurs between a phase with effectively non-interacting particles (``vacuum'')
and a phase of strongly coupled bosons that form Bose-Einstein condensate (BEC) in dimension $d$.
The dual $d+1$-dimensional disorder-driven phase transition, which we predict here, occurs, respectively, between a phase with effectively vanishing disorder 
and a strongly disordered phase. This disorder-driven transition manifests itself in the critical behaviour of observables 
such as the density of states and transport coefficients in the system.
In contrast with the previously studied non-Anderson disorder-driven transitions~\cite{Syzranov:review}
the instability predicted here is described exactly by the 
RG Eq.~\eqref{RGflowCoupling}, which allows for an exact determination of the correlation-length critical exponent $\nu=1/(d-\alpha)$.

\begin{figure}[t!]
	\centering
	\includegraphics[width=0.6\linewidth]{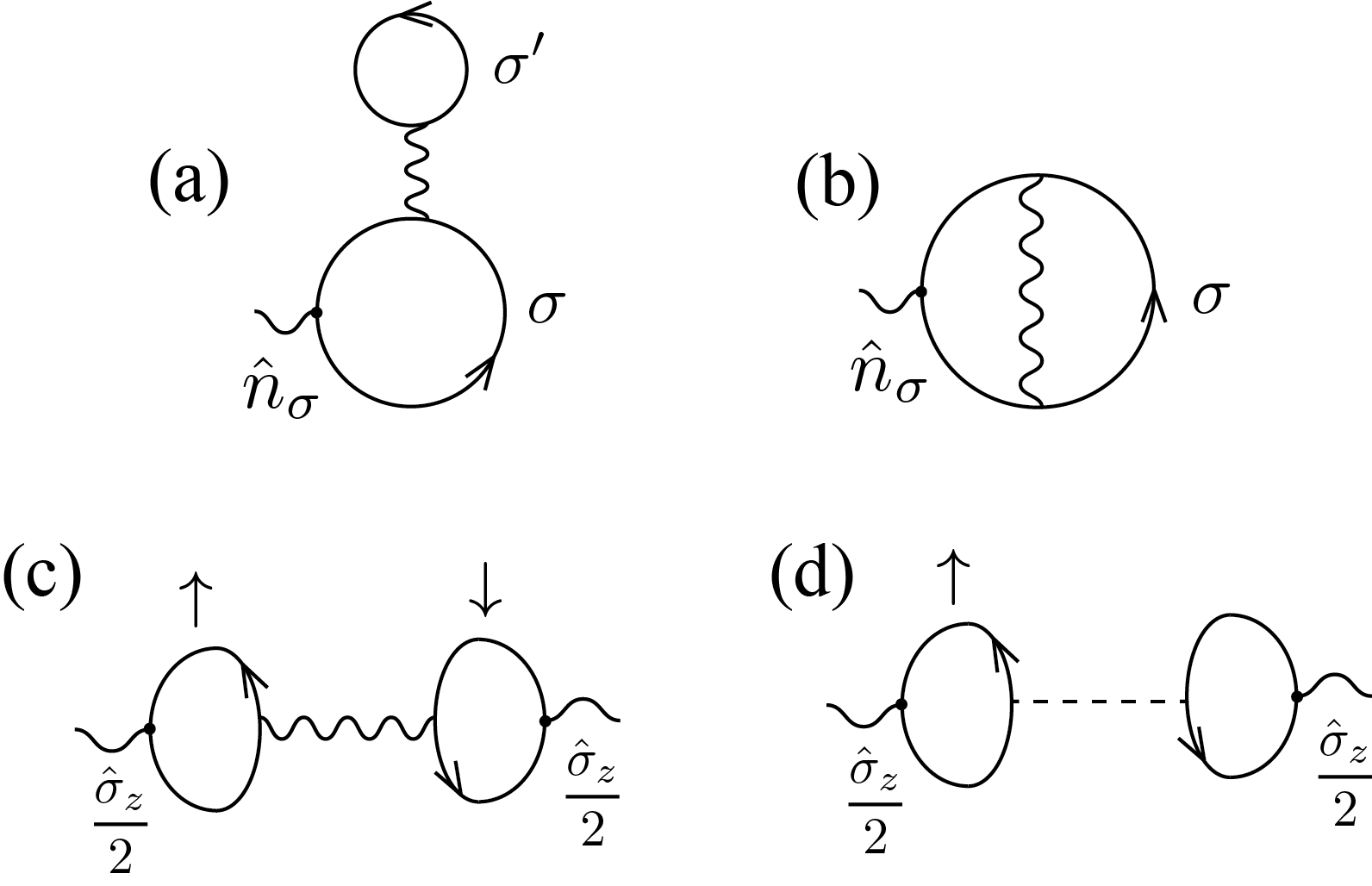}
	\caption{\label{fig:diagramsdot}
		Diagrams for correlations in the one-site Hubbard model and a disordered wire. Diagrams (a) and (b) describe modifications to the number $n_\sigma$ of the particles with spin $\sigma$, but do not affect the correlations between $n_\uparrow$ and $n_\downarrow$ mimicked by diagram (c) to the first order in the interaction $g$. Diagram (d) describes correlations in the dual non-interacting disordered system.}
\end{figure}

{\it Low-order correlations in electronic systems: a quantum dot and a 1D disordered wire.}
Above, we described classes of interacting and disordered systems that are associated to each other by the 
derived duality mapping.
In what immediately follows, we demonstrate that the mapping can generically be applied to correlators
of observables, e.g. electron densities, if they are considered to the leading order 
in interactions, in which they are not affected by screening and the Hartree contributions.
To illustrate this, we consider a one-site Hubbard model (quantum dot) described by the Hamiltonian
\begin{align}
	\cH_{dot}=\xi \hn_\uparrow +\xi \hn_\downarrow -g\hn_\uparrow\hn_\downarrow,
	\label{Hdot}
\end{align}
where $\xi$ is a constant and $n_\uparrow$ and $n_\downarrow$ are the numbers of the electrons in the ``spin up'' and ``spin down'' states.
This quantum dot is dual to a system of a 1D particle in a random potential $u(x)$, with the Hamiltonian given by
\begin{align}
	\hh_{wire}=\sum_{i=\uparrow,\downarrow}\hPsi_i(x)\left[\xi\hsigma_z-i\hsigma_y\partial_x+u(x)\hsigma_z\right]\hPsi_i(x).
	\label{Hwires}
\end{align}
In general, perturbative contributions with loops are not negligible for electrons in a quantum dot described by the 
Hamiltonian~\eqref{Hdot}. For example, the Hartree contribution to the number of electrons $n_\sigma$
with spin $\sigma$ in Fig.~\ref{fig:diagramsdot}a matches the value of diagram~\ref{fig:diagramsdot}b. These contributions, however, do not affect, to the first order in the coupling $g$, the correlator
\begin{equation}
	K=\langle\hat{n}_{\uparrow}\hat{n}_{\downarrow}\rangle-\langle\hat{n}_{\uparrow}\rangle\langle\hat{n}_{\downarrow}\rangle
	=\frac{g}{T}\left[\frac{e^{\xi/T}}{(1+e^{\xi/T})^{2}}\right]^{2}+{\cal O}\left(g^2\right)
	\label{qdotn}
\end{equation}
of the numbers $n_\uparrow$ and $n_\downarrow$ of fermions with different spins computed in Supplemental Material~\cite{SupplementalInformation}.
The value of the correlator \eqref{qdotn} matches, to the leading order in $g$, the correlator
\begin{equation}
	K_{\text{dis}}=
	\langle{\rho}_{s\uparrow}{\rho}_{s\downarrow}\rangle_{\text{dis}}-
	\langle{\rho}_{s\uparrow}\rangle_{\text{dis}}\langle{\rho}_{s\downarrow}\rangle_{\text{dis}}
	\label{qdotnDisorder}
\end{equation}
of the operators $\rho_{s}$ given by Eq.~\eqref{RhoDefinition}. The value of the correlator \eqref{qdotnDisorder} is computed independently for 
a disordered wire in Supplemental Material~\cite{SupplementalInformation}.
The matching of the correlators \eqref{qdotn} and \eqref{qdotnDisorder} to the leading order in $g$ 
reflects the duality between interactions and quenched disorder discussed in this paper.

{\it Conclusion and outlook.} In summary, we have demonstrated the equivalence of a class of disorder-free interacting systems
to non-interacting disordered systems. The interacting systems that allow for this mapping include dilute quantum gases of trapped ultracold particles and nodal semimetals, in which the screening of the interactions is suppressed due to the vanishing DoS near the node. The mapping may also be applied to interacting systems with large Fermi surfaces if the screening
has no qualitative effect on the observable quantities, e.g. does not change the universality class of a phase transition.

Furthermore, the mapping allows to predict new phenomena, e.g. phase transitions, in disordered and interacting systems dual to
previously known phenomena.
The duality can also be applied to map interacting systems to non-Hermitian disordered systems (by, e.g., applying the 
version of the mapping to attractive interactions to systems with repulsive interactions or by skipping
the ``Hermitisation'' step in the described construction of the mapping). It can thus be used to explore and describe
phase transitions and other phenomena in systems described by non-Hermitian Hamiltoanians ~\cite{Konotop:nonHermitianHamiltonians,YoshidaHatsugai:nonHermitianReview,Varguet:nonHermitianReview,ShenFu:nonHermitianTopology,HatanoNelson:VortexNonHermitian},
which we leave for future studies.

Using this mapping, we predict novel non-Anderson disorder driven transitions, such as the disorder-driven transition in 
nodal semimetals dual to the BEC-vacuum transition known previously for interacting bosonic systems. We confirm the existence 
of such a transition by rigorous microscopic calculations. 
Based on the established duality we also expect a novel disorder-driven transition in 3D nodal-line semimetals and new
interaction-driven transitions in systems with power-law dispersions.

Other questions that remain to be investigated is the role of the non-perturbative (instantonic) effects on the predicted 
phenomena as well as the possibility of spontaneously generated relevant operators that may change the criticality
at the phase transitions~\cite{SbierskiSyzranov:Thouless}.

{\it Acknowledgments.} We are grateful to B.~Sbierski, L.~Radzihovsky and S.~Zhu for insightful discussions and comments on the manuscript.
We have also benefited from discussion with Ya.~Rodionov at the early stages of the project.


\bibliography{references}


\onecolumngrid
\vspace{2cm}

\cleardoublepage

\renewcommand{\theequation}{S\arabic{equation}}
\renewcommand{\thefigure}{S\arabic{figure}}
\renewcommand{\thetable}{S\arabic{table}}
\renewcommand{\thetable}{S\arabic{table}}
\renewcommand{\bibnumfmt}[1]{[S#1]}
\renewcommand{\citenumfont}[1]{S#1}

\setcounter{equation}{0}
\setcounter{figure}{0}
\setcounter{enumiv}{0}

\pagestyle{empty}




%


\begin{center}
	\textbf{\large Supplemental Material for \\
		``Interacting systems equivalent to
		non-interacting systems with quenched disorder''
	}
	\\
	Shijun Sun and Sergey Syzranov
\end{center}

\section{Details of the mapping to all orders of the perturbation theory}

\label{Sec:Mapping_Details}

In this section, we provide the details of the duality between disordered non-interacting and interacting disorder-free systems.
Dual quantities are summarised in Table~\ref{Table}.
\begin{table}[ht!]
	\begin{center}
		\begin{tabular}{l| c| c}
			\toprule 
			\, & {Interacting model} & {Disordered model}\\
			\hline 
			Coordinates & $(\tau,\mathbf{r})$ & $(\mathrm{r}_{d+1},\mathbf{r})$ \\
			\hline
			Temperature/size & $T$ & $1/\ell_{d+1}$ \\
			\hline
			\makecell[l]{Coupling to \\ interactions/disorder}
			& $\hPsi^\dagger\hPsi\phi$ & $\hpsi^\dagger\hsigma_z\hpsi u$ \\
			\hline
			Observables & $\hat{n}$ (density) & ${\rho}_{s}$, Eq.~\eqref{RhoDefinition}\\
			\bottomrule
		\end{tabular}
	\end{center}
	\caption{\label{Table} Correspondence between quantities in the interacting disorder-free and non-interacting disordered systems}
\end{table}
As an example of an observable quantity in the interacting system, we use the density of particles, dual to the quantity $\rho_{s}$ given by Eq.~\eqref{RhoDefinition}
in the disordered system. The density of particles and the quantity $\rho_s$ are represented by 
sets of diagrams in Figs.~\ref{fig:maindiagrams}a and \ref{fig:maindiagrams}d.
Basic elements of the diagrammatic technique are shown in Fig.~\eqref{fig:elementsDiagrammatics}.
We describe first a generic diagram for an interacting system and demonstrate its equivalence to the corresponding diagram for the non-interacting disordered system.
Then we provide explicit expressions for several lowest-order diagrams in both systems.
 
The value of each diagram with $N$ interaction propagators contributing to the density of particles in the interacting system is given by
\begin{align}
    (-1)^{N+F}\frac{T^{N+1}}{V^{N+1}}\sum_{\omega,\bp}\frac{1}{(i\omega_0-\xi_{\bp_0})^2}
	\frac{1}{i\omega_1-\xi_{\bp_1}}\ldots\frac{1}{i\omega_{2N-1}-\xi_{\bp_{2N-1}}}\, D(\Omega_{1},\bP_{1})\ldots D(\Omega_{N},\bP_{N}),
	\label{AppendixInteractingDiagramExpression}
\end{align}
where $F=1$ for fermionic particles and $F=0$ for bosonic particles; $D(\Omega_{i},\bP_{i})$ is the interaction propagators which depends on the bosonic (fermionic) Matsubara
 frequency $\Omega_i=2\pi T n_i$ [$\Omega_i=\pi T (2n_i+1)$] and momentum $\bP_i$ and is the Fourier-transform of the interaction propagator
\begin{align}
	D(\br,\tau;\br^\prime,\tau^\prime)=-\left<T_\tau \hat\phi(\br,\tau)\hat\phi(\br^\prime,\tau^\prime)\right>
\end{align} 
in the coordinate and Matsubara-time representation, where $\hat\phi$ are the bosonic fields corresponding to the interaction between the particles.
The summation $\sum_{\omega,\bp}\ldots$ in Eq.~\eqref{AppendixInteractingDiagramExpression} may be carried out over any $N+1$ independent frequencies and momenta, with the other frequency and momenta of the particle and interaction propagators determined from the energy and momentum 
conservation laws in the diagram. We assume the convergence of the sum for each diagram.

In the case of short-ranged unscreened interactions, each interaction propagator is frequency- and momentum-independent and may be replaced by a constant,
$D(\Omega_{i},\bP_{i})\rightarrow -g$, given by Eq.~\eqref{Coupling}. In this section, we do not assume any specific form of the interaction propagator $D(\Omega,\bP)$.

Because the bosonic propagator $D(\Omega_{i},\bP_{i})$ is an even function of the Matsubara frequency $\Omega_i$, each summation
with respect to Matsubara frequencies $\omega$ in Eq.~\eqref{AppendixInteractingDiagramExpression} can be replaced with two summations with respect to $\omega$
and $-\omega$, $\sum_{\omega}\ldots=\frac{1}{2}\sum_{\omega}\ldots+\frac{1}{2}\sum_{-\omega}\ldots$, which gives
\begin{align}
    (-1)^{N+F}\frac{T^{N+1}}{2 V^{N+1}}\sum_{I=0,1}\sum_{\omega,\bp}\frac{1}{\left[(-1)^{I}i\omega_0 -\xi_{\bp_0}\right]^2}
	\frac{1}{(-1)^{I}i\omega_1-\xi_{\bp_1}}\ldots\frac{1}{(-1)^{I}i\omega_{2N-1}-\xi_{\bp_{2N-1}}}\, D(\Omega_{1},\bP_{1})\ldots D(\Omega_{N},\bP_{N}).
	\label{AppendixInteractingDiagramFreqInv}
\end{align}
Below, we compare the expression \eqref{AppendixInteractingDiagramFreqInv} for the $N$-th-order diagram for an interacting disorder-free system to 
the value of a similar diagram in the equivalent disordered non-interacting system.

In what immediately follows, we assume that the equivalent disordered system described by the Hamiltonian~\eqref{HamiltonianDisordered}
has the volume $V\ell_{d+1}$ in the dimension $d+1$, 
where $\ell_{d+1}$ is its length along one dimension and $V$ is the cross section in the remaining $d$ dimensions.
The topologically equivalent $N$-th order diagram 
is given by
\begin{align}
	\frac{(-1)^{N+F}}{\widetilde{V}^{N+1} \ell_{d+1}^{N+1}}\sum_{\bp,k}\mathrm{Tr}
	\left[\hsigma_z\frac{1}{-k_0\hsigma_y-\xi_{\bp_{0}}\hsigma_z} \frac{\hsigma_z}{2} \frac{1}{-k_0\hsigma_y-\xi_{\bp_{0}}\hsigma_z} \hsigma_z
	\ldots\hsigma_z\frac{1}{-k_{2N-1}\hsigma_y-\xi_{\bp_{2N-1}}\hsigma_z}
	\right]\widetilde{D}(K_{1},\bP_{1})\ldots \widetilde{D}(K_{N},\bP_{N}),
	\label{AppendixDisorderDiagramExpression}
\end{align}
where $(\bp_i,k_i)$ is a $d+1$-dimensional momentum; $i=0,1,\ldots 2N-1$;
$\mathrm{Tr}\ldots$ is taken with respect to the pseudospin degrees of freedom; (anti-)periodic boundary conditions have to be chosen along the $d+1$-st dimension for (fermionic) bosonic particles in the interacting 
system;  $-\widetilde{D}(K_i,\bP_i)$ is the ``impurity line''~\cite{AGD}, the Fourier-transform 
of the correlator
\begin{align}
	-\widetilde{D}(\brho-\brho^\prime)=\left<u(\brho)u(\brho^\prime)\right>_{\text{dis}}
\end{align}
of the random potential $u(\brho)$. Here, in accordance with the common convention, the impurity line (cf. Fig.~\ref{fig:elementsDiagrammatics}) is defined to be positive for a real random potential.
 Similarly to the case of the diagram for the interacting system, the summation in Eq.~\eqref{AppendixDisorderDiagramExpression} may be carried out over any
$N+1$ independent momenta in the dimension $d+1$, while the other momenta of the particle and disorder propagators are determined from the law of momentum conservation.
In Eq.~\ref{AppendixDisorderDiagramExpression}, we took into account that the the quantity $\rho_s$, to which the respective diagram contributes, corresponds to the $\hsigma_z/2$ vertex 
and to the particle propagator
\begin{align}
	\left(-k_i\hsigma_y-\xi_{\bp_{i}}\hsigma_z\right)^{-1}=\frac{1}{2}\left[G^A(k_i,\bp_i,E=0)+G^R(k_i,\bp_i,E=0)\right],
\end{align}
where $G^A$ and $G^R$ are the advanced and retarded Green's functions of a free particle.

Equation~\eqref{AppendixDisorderDiagramExpression} gives
\begin{align}
	\frac{(-1)^{N+F}}{2 \widetilde{V}^{N+1} \ell_{d+1}^{N+1}}\sum_{\bp,k}\mathrm{Tr}
	\left[\frac{1}{i k_0\hsigma_x-\xi_{\bp_{0}}\mathbbm{1}_{2\times2}}  \frac{1}{i k_0\hsigma_x-\xi_{\bp_{0}}\mathbbm{1}_{2\times2}} 
	\ldots\frac{1}{i k_{2N-1}\hsigma_x-\xi_{\bp_{2N-1}}\mathbbm{1}_{2\times2}}
	\right]\widetilde{D}(K_{1},\bP_{1})\ldots \widetilde{D}(K_{N},\bP_{N}),
	\label{IntermediteUnimportantFormula}
\end{align}
where $\mathbbm{1}_{2\times2}$ is the identity matrix in the pseudospin space.
Because the eigenvalues of the operator $\hsigma_{x}$ are given by $(-1)^I$ with $I=0,1$,
Eq.~\eqref{IntermediteUnimportantFormula} can
be rewritten as 
\begin{align}
	\frac{(-1)^{N+F}}{2 \widetilde{V}^{N+1} \ell_{d+1}^{N+1}}\sum_{I=0,1}\sum_{\bp,k}
	\left[\frac{1}{i(-1)^{I} k_0-\xi_{\bp_{0}}}  \frac{1}{i(-1)^{I} k_0-\xi_{\bp_{0}}} 
	\ldots\frac{1}{i(-1)^{I} k_{2N-1}-\xi_{\bp_{2N-1}}}
	\right]\widetilde{D}(K_{1},\bP_{1})\ldots \widetilde{D}(K_{N},\bP_{N}).
	\label{AppendixDisorderDiagramEigenbasis}
\end{align}

Equations~\eqref{AppendixInteractingDiagramFreqInv} and \eqref{AppendixDisorderDiagramEigenbasis} for the diagrams for, respectively,
the interacting disorder-free and non-interacting disordered systems are identical to each other so long as $\ell_{d+1}=1/T$ and
the Matsubara frequencies $\omega_i$ in Eq.~\eqref{AppendixInteractingDiagramFreqInv} match 
the values of the momenta
$k_i$ in Eq.~\eqref{AppendixDisorderDiagramEigenbasis}. 
%
The latter condition, with $k_i=2\pi T n_i$ ($k_i=2\pi T n_i+\pi T$) and integer $n_i$,
is satisfied if (anti-)periodic boundary conditions are imposed on the disordered system in the case of a (fermionic) bosonic interacting system.

{\it Summary.} In summary, we have established the correspondence, to all order of the perturbation theory, between
observables in a $d$-dimensional bosonic (fermionic) interacting disorder-free system at temperature $T$
and a dual $d+1$-dimensional non-interacting disordered system of length $\ell_{d+1}=1/T$
 with (anti-)periodic boundary conditions along the $d+1$-st dimension. 
We focussed on the observable quantities
\begin{equation}
	\langle\hn(\br)\rangle=\langle\hat{\rho}_{s}(\brho)\rangle_\text{dis},
\end{equation}
where $\hn$ is the density of particles in the interacting system and the operator $\rho_s$ in the disordered system is defined by Eq.~\eqref{RhoDefinition}.
The established equivalence applies, however, to other observables, such as currents and spin/valley degrees of freedom and their correlators.
To further illustrate the discussed duality, we consider below the zeroth and first order diagrams contributing to $\langle\hn(\br)\rangle$
and $\langle\hat{\rho}_{s}(\brho)\rangle_\text{dis}$ explicitly.

\subsection{Zeroth order} 
The concentration of particles at the zeroth order is given by
\begin{align}
    \langle\hn^{(0)}(\br)\rangle=\frac{T}{V}{\sum_{\omega,\bp}}^\prime
    \frac{(-1)^{F}}{i\omega-\xi_{\bp}}=\frac{1}{V}\sum_{\bp}\frac{1}{\exp\left(\xi_{\bp}/T\right)\mp 1},
    \label{AppendixInteractionZeroth}
\end{align}
where $\sum^\prime$ is our convention for the regularised sum over Matsubara frequencies (which amounts to, e.g., infinitesimal phase corrections to the frequencies~\cite{AltlandSimons:book} $i\omega\rightarrow i\omega e^{-i\omega\delta}$), ensuring that the sum of a Matsubara Green's function over frequencies gives the Bose (Fermi) distribution function for bosonic (fermionic) frequencies.

For the disordered system, the zeroth-order contribution to the dual observable is given by
\begin{align}
    \langle{\rho}_{s}^{(0)}(\brho)\rangle_{\textrm{dis}}=
    \frac{(-1)^{F}}{\widetilde{V}\ell_{d+1}}{\sum_{\bp,k}}^\prime\mathrm{Tr}\left[\frac{\hat{\sigma}_{z}}{2}\frac{1}{-k\hat{\sigma}_{y}-\xi_{\bp}\hat{\sigma}_{z}}\right]=
    \frac{(-1)^{F}}{\widetilde{V}\ell_{d+1}}{\sum_{\bp,k}}^\prime\frac{-\xi_{\bp}}{k^{2}+\xi_{\bp}^{2}}=
    \frac{1}{\widetilde{V}}{\sum_{\bp}}^\prime\frac{1}{\exp\left(\xi_{\bp}\ell_{d+1}\right)\pm 1},
    \label{AppendixDisorderZeroth}
\end{align}
where ``+'' and ``-'' correspond, respectively, to periodic and antiperiodic boundary conditions along the $d+1$-st dimension, 
resulting in the quantised values $k=2\pi\ell_{d+1}^{-1}n$ and $k=\pi\ell_{d+1}^{-1}(2n+1)$ of the momentum $k$.
Equations~\eqref{AppendixInteractionZeroth} and \eqref{AppendixDisorderZeroth} are precisely equivalent for $\ell_{d+1}=1/T$, in accordance with the duality transformation derived in this paper.


\subsection{First order}

\begin{figure}[t!]
	\centering
	\includegraphics[width=0.5\linewidth]{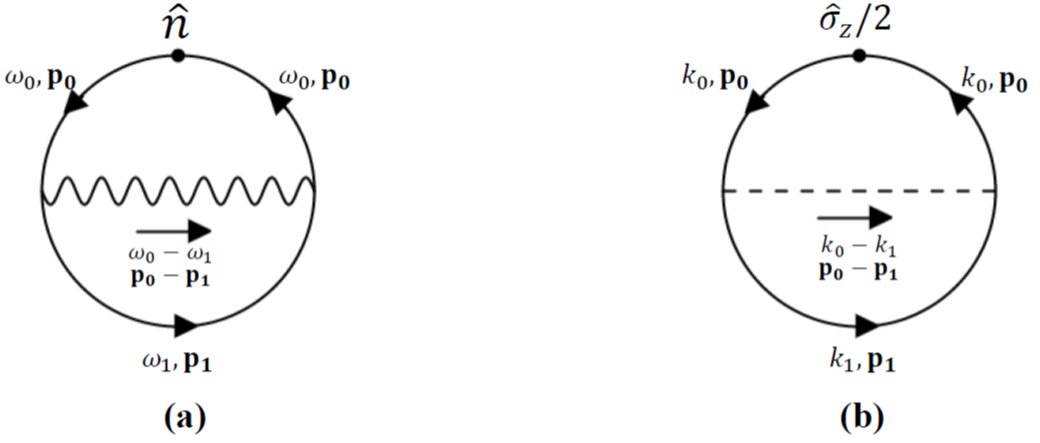}
	\caption{\label{fig:diagram1loop} First-order diagrams for the density $\hn$ in interacting disorder-free ($a$) and the operator $\rho_s$ in non-interacting 
	disordered ($b$) systems. 
	}
\end{figure}
Fig.~\ref{fig:diagram1loop}a shows the first-order correction to $\langle\hn\rangle$. This diagram contributes
\begin{align}
    (-1)^{F+1}\frac{T^{2}}{V^{2}}\sum_{\omega_{0},\omega_{1},\bp_{0},\bp_{1}}\frac{1}{(i\omega_0-\xi_{\bp_0})^2}
	\frac{1}{i\omega_1-\xi_{\bp_1}}D(\omega_{0}-\omega_{1},\bp_{0}-\bp_{1}).
	\label{Appendix1loopInteraction}
\end{align}
Again, because the bosonic propagator is even under the inversion of Matsubara frequency, $D(\omega_{0}-\omega_{1},\bp_{0}-\bp_{1})=D(-\omega_{0}+\omega_{1},\bp_{0}-\bp_{1})$, the sum with respect to the frequencies in Eq.\eqref{Appendix1loopInteraction} is equivalent to two sums with respect to $\omega_0,\omega_1$ and $-\omega_0,-\omega_1$, $\sum_{\omega_{0},\omega_{1}}\ldots=\frac{1}{2}\sum_{\omega_{0},\omega_{1}}\ldots+\frac{1}{2}\sum_{-\omega_{0},-\omega_{1}}\ldots$. Therefore, Eq.\eqref{Appendix1loopInteraction} becomes
\begin{align}
    (-1)^{F+1}\frac{T^{2}}{2V^{2}}\sum_{\omega_{0},\omega_{1},\bp_{0},\bp_{1}}\left[\frac{1}{(i\omega_0-\xi_{\bp_0})^2}
	\frac{1}{i\omega_1-\xi_{\bp_1}}+\frac{1}{(-i\omega_0-\xi_{\bp_0})^2}
	\frac{1}{-i\omega_1-\xi_{\bp_1}}\right]D(\omega_{0}-\omega_{1},\bp_{0}-\bp_{1}).
	\label{Appendix1loopInteraction2}
\end{align}
 
The corresponding diagram for the non-interacting disordered system is shown in Fig.~\ref{fig:diagram1loop}b and is given by
\begin{align}
    \frac{(-1)^{F}}{\widetilde{V}^{2} \ell_{d+1}^{2}}\sum_{\bp_{0},\bp_{1},k_{0},k_{1}}\mathrm{Tr}
	\left[
	 \hsigma_z\frac{1}{-k_0\hsigma_y-\xi_{\bp_{0}}\hsigma_z} \frac{\hsigma_z}{2} \frac{1}{-k_0\hsigma_y-\xi_{\bp_{0}}\hsigma_z} \hsigma_z \frac{1}{-k_{1}\hsigma_y-\xi_{\bp_{1}}\hsigma_z}
	\right]\left[-\widetilde{D}(k_{0}-k_{1},\bp_{0}-\bp_{1})\right]\nonumber\\
	=\frac{(-1)^{F+1}}{2\widetilde{V}^{2} \ell_{d+1}^{2}}\sum_{\bp_{0},\bp_{1},k_{0},k_{1}}\mathrm{Tr}
	\left[\frac{1}{i k_0\hsigma_x-\xi_{\bp_{0}}\mathbbm{1}_{2\times2}}  \frac{1}{i k_0\hsigma_x-\xi_{\bp_{0}}\mathbbm{1}_{2\times2}} \frac{1}{i k_1\hsigma_x-\xi_{\bp_{0}}\mathbbm{1}_{2\times2}}
	\right]\widetilde{D}(k_{0}-k_{1},\bp_{0}-\bp_{1}).
\end{align}
Taking the trace with respect to the eigenvalues of $\hsigma_x$ gives
\begin{align}
    \frac{(-1)^{F+1}}{2\widetilde{V}^{2} \ell_{d+1}^{2}}\sum_{\bp_{0},\bp_{1},k_{0},k_{1}}
	\left(\frac{1}{i k_0-\xi_{\bp_{0}}}  \frac{1}{i k_0-\xi_{\bp_{0}}} \frac{1}{i k_1-\xi_{\bp_{0}}}
	+\frac{1}{-i k_0-\xi_{\bp_{0}}}  \frac{1}{-i k_0-\xi_{\bp_{0}}} \frac{1}{-i k_1-\xi_{\bp_{0}}}\right)\widetilde{D}(k_{0}-k_{1},\bp_{0}-\bp_{1}).
	\label{Appendix1loopDisorder}
\end{align}
Since the values of $\omega_{i}$ and $k_{i}$ match, due to the choice of the boundary conditions, and  $\ell_{d+1}=1/T$,
Eqs.~\eqref{Appendix1loopInteraction2} and \eqref{Appendix1loopDisorder} are identical.



\section{Renormalisation-group approach to interacting gases and high-dimensional semimetals}

\label{Sec:RGdetails}

%
%
%
%
%
%
%
%

In this section, we describe the renormalisation of interactions in gases of particles with the power-law dispersion $\xi_{\bk}\propto k^\alpha$
and the renormalisation of disorder in the dual class of systems, i.e. semimetals with the dispersion $\xi_\bk\hsigma_z+k_{d+1}\hsigma_y$.
We demonstrate that these renormalisations are described by the same RG flow equation~\eqref{RGflowCoupling}, which illustrates that these systems exhibit interaction-driven (disorder-driven) phase transitions in the same universality class.
\begin{figure}[th!]
	\centering
	\includegraphics[width=0.6\linewidth]{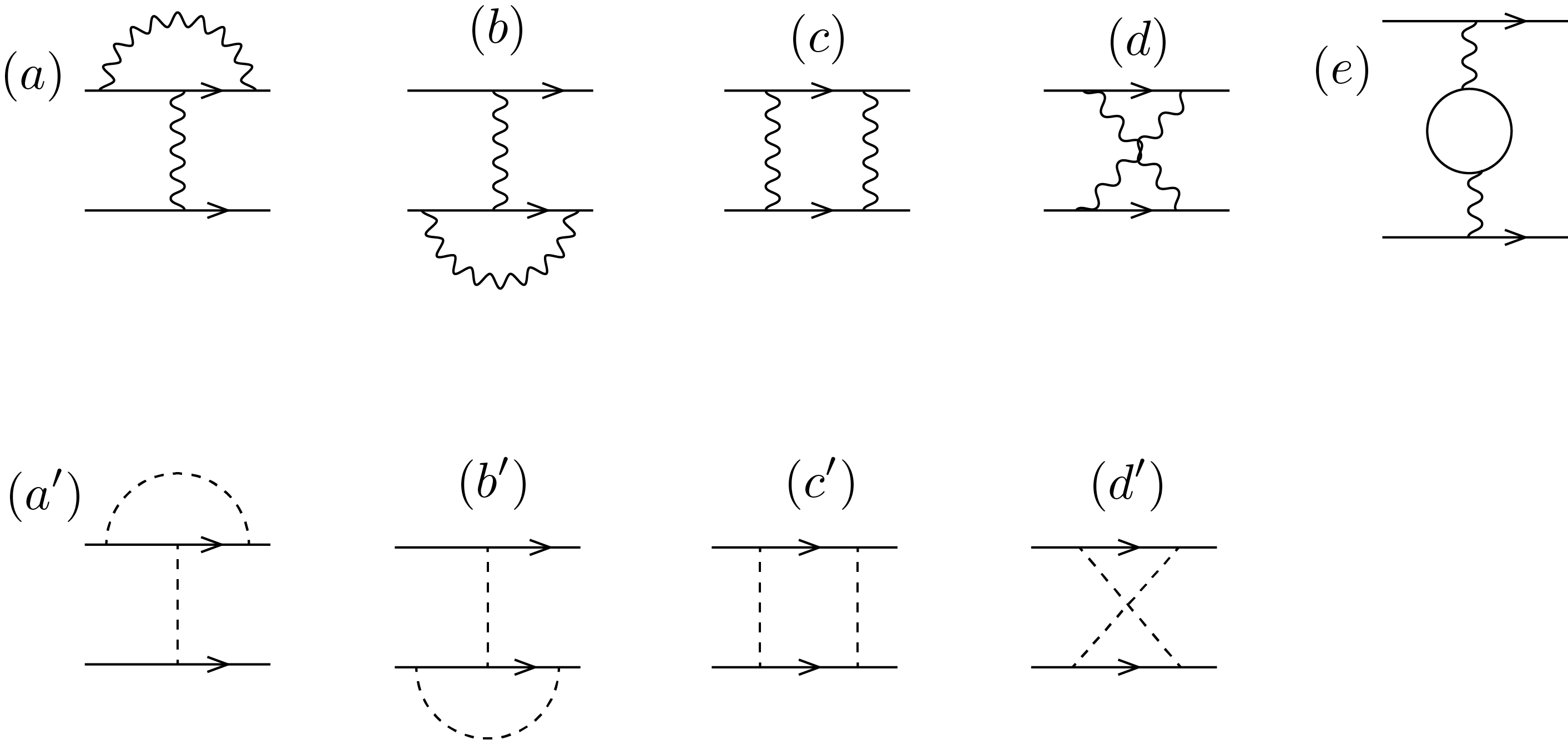}
	\caption{\label{fig:diagramsrg} Diagrams for the renormalisation of the coupling constants in an interacting disorder-free ($a-e$) and non-interacting 
		disordered ($a^\prime-d^\prime$) systems. }
\end{figure}

The RG procedure for the interacting system involves repeatedly integrating out shells of largest momenta and frequencies,
\begin{subequations}
\begin{align}
	K e^{-l}\ll|\bk|\ll K,
	\label{MomentumIntervalRG}
	\\
	\left|\xi_{K e^{-l}}\right|\ll \omega\ll\left|\xi_K\right|,
	\label{FrequencyIntervalRG}
\end{align}
\end{subequations}
and renormalising the properties of the systems perturbatively in the coupling constant $g$.
The details of the cutoff procedure are not important in the one-loop approximation for the dimension $d$ near the critical dimension $d_c=\alpha$.
 The diagrams for the one-loop renormalisation of the interaction propagator are shown in 
Fig.~\ref{fig:diagramsrg}$a$-$e$. 
When evaluating them, it is sufficient to set all external incoming and outgoing frequencies and momenta to zero and sum/integrate only with respect to intermediate frequencies and momenta.
The main contribution comes from diagram~\ref{fig:diagramsrg}$c$:
\begin{align}
	[\ref{fig:diagramsrg}c]=g^2 T\sum_{i\omega}\int_\bk \frac{1}{i\omega-\hxi_\bk}\otimes\frac{1}{-i\omega-\hxi_{-\bk}},
	\label{ImportantDiagramRG}
\end{align}
where the frequency summation and integration with respect to the momentum $\bk$ are carried out over the intervals \eqref{MomentumIntervalRG}-\eqref{FrequencyIntervalRG}; $\int_\bk\ldots =\int \frac{d^d\bk}{(2\pi)^d}\ldots$;
the dispersion $\xi_\bk\propto k^\alpha$ has the power dependence on the momentum $\bk$, but may also have additional structure in the valley or spin space; $\otimes$ is the product of the spin/valley subspaces corresponding to the top and bottom propagators in Fig.~\eqref{fig:diagramsrg}$c$.

We consider the case of  large ultraviolet momentum cutoffs $K$ and $K e^{-l}$, corresponding to the kinetic energies significantly exceeding the temperature $T$. This allows us to replace the summation with respect to frequencies in Eq.~\eqref{ImportantDiagramRG} by integration, $T\sum_{i\omega}\ldots\rightarrow\int\frac{d\omega}{2\pi}\ldots$. For a scalar dispersion 
\begin{align}
	\xi_\bk=|\bk|^\alpha,
\end{align}
which has no valley and spin structure, the renormalised interaction propagator also has a trivial structure ($\propto \holOne\otimes\holOne$) in the spin/valley space, and the value of diagram~\ref{fig:diagramsrg}$c$ is given by
\begin{align}
	[\ref{fig:diagramsrg}c]=\frac{g^2 S_d K^{d-\alpha}}{2(2\pi)^d}\frac{1-e^{-(d-\alpha)l}}{d-\alpha}.
	\label{DiagramTakenIntoAccount}
\end{align}
All the other contributions shown in Fig.~\ref{fig:diagramsrg} may be estimated as
\begin{align}
	[\ref{fig:diagramsrg}a]\sim[\ref{fig:diagramsrg}b]\sim[\ref{fig:diagramsrg}d]\sim[\ref{fig:diagramsrg}e]
	\sim \frac{g^2 S_d K^{d-\alpha}}{2(2\pi)^d}
\end{align}
and are suppressed for the dimensions $d$ close to the critical dimension $d_c=\alpha$. This leads to the RG flow equation for the coupling $g$ given by
\begin{align}
	\partial_l g=\frac{ S_d K^{d-\alpha}}{2(2\pi)^d}g^2.
\end{align}
Introducing the dimensionless coupling constant 
\begin{align}
	\gamma=\frac{S_d}{2(2\pi)^d}gK^{d-\alpha}
\end{align}
gives the one-loop RG flow equation
\begin{align}
	\partial_l\gamma=(\alpha-d)\gamma+\gamma^2.
	\label{RGequation}
\end{align}

\begin{figure}[h!]
	\centering
	\includegraphics[width=0.3\linewidth]{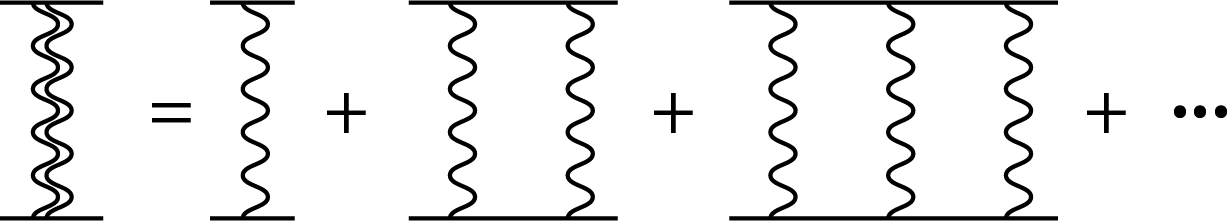}
	\caption{\label{fig:ladder} Ladder diagrams for the renormalisation of the interaction between bosonic particles in vacuum.}	
\end{figure}

It is possible to show that the RG flow equation~\eqref{RGequation} is exact, i.e. applies beyond the one-loop approximation.
It has been noticed in Ref.~\onlinecite{Uzunov:BCSRG} that the renormalised contact interaction between quadratically dispersive bosonic particles is given by the ladder diagrams shown in Fig.~\eqref{fig:ladder} and is, therefore, corresponding to the solution of the RG equation~\eqref{RGequation}.
This result can be straightforwardly generalised to the case of the power-law dispersion $\xi_\bk=k^\alpha$ considered here.
The RG flow is terminated at the value of the ultraviolet cutoff $K$ equal to the inverse size $L^{-1}$ or a characteristic momentum scale corresponding to the renormalised kinetic energy on the order of the temperature $T$ or the chemical potential $\mu$.

The diagrams for the renormalisation in the dual disordered non-interacting system, described by the Hamiltonian~\eqref{HamiltonianDisordered}, are shown in Figure~\ref{fig:diagramsrg}$a^\prime-d^\prime$. They are topologically equivalent to diagrams~\ref{fig:diagramsrg}$a-d$ and do not include a diagram with a closed loop of particle propagators. In the diagrammatic technique for the disordered systems~\cite{AGD}, contributions with loops are absent by construction.
Although such loops are present for the interacting systems we consider, their contribution is suppressed due to the suppressed density of states at the chemical potential assumed in this paper.

The main contribution to the renormalisation of the coupling $g$ in the disordered system comes from diagram~\ref{fig:diagramsrg}$c^\prime$.
While the other contributions to the renormalisation are suppressed, it is convenient to evaluate together diagrams~\ref{fig:diagramsrg}$c^\prime$ and \ref{fig:diagramsrg}$d^\prime$:
\begin{align}
	\ref{fig:diagramsrg}c^\prime+\ref{fig:diagramsrg}d^\prime
	&=g^2\int_\bp \int_{p_{d+1}}\hsigma_{z}\left(\frac{1}{\xi_\bp\hsigma_{z}+p_{d+1}\hsigma_y}+\frac{1}{\xi_{-\bp}\hsigma_{z}-p_{d+1}\hsigma_y}\right)\hsigma_z
	\otimes \hsigma_z\frac{1}{\xi_\bp\hsigma_{z}+p_{d+1}\hsigma_y}\hsigma_z
	\nonumber\\
	&=g^2\int_\bp \int_{p_{d+1}} \frac{2\xi_\bp\hsigma_{z}}{\xi_\bp^2+p_{d+1}^2}
	\otimes\frac{\hsigma_z\left(\xi_\bp\hsigma_z+p_{d+1}\hsigma_y\right)\hsigma_z}{\xi_\bp^2+p_{d+1}^2}
	=g^2\int_\bp \int_{p_{d+1}}\frac{2\xi_\bp^2}{\left[\xi_\bp^2+p_{d+1}^2\right]^2}\hsigma_z\otimes\hsigma_z
	\nonumber\\
	&\approx\frac{g^2 S_d K^{d-\alpha}}{2(2\pi)^d}\frac{1-e^{-(d-\alpha)l}}{d-\alpha}.
\end{align}
Similarly to the case of interacting systems, it is possible to demonstrate that the contributions of the other diagrams to the renormalisation of the coupling $g$ are suppressed, and the flow of the coupling is again described by Eq.~\eqref{RGequation}.
Identical RG flows for the coupling in the cases of interacting disorder-free and non-interacting  disordered systems illustrate the equivalence between the two classes of systems discussed in this paper.

Both of these classes of systems display 
transitions at the critical value of the dimensionless coupling $\gamma_c=d-\alpha$ between the phases with irrelevant interactions (disorder),
for subcritical coupling, and relevant interactions (disorder), for supercritical interactions (disorder).
For the example of the disordered-driven transition considered here, the universality classes of the dual transitions match exactly,
owing to the absence of screening of the interactions in the vacuum phase of the interacting system at zero temperature and chemical potential.
If the interacting system has a small chemical potential $\mu$ or temperature $T$, the mapping becomes approximate. 
At small values of $K_\text{ir}=\left[\min(|\mu|,T)\right]^\frac{1}{\alpha}\ll K_0$, however, the mapping will still be accurate, as is clear from  
comparing the value of the loop diagram $[\ref{fig:diagramsrg}e]\sim g^2\frac{S_d}{(2\alpha\pi)^d} K_\text{ir}^{d-\alpha}$
to the contribution of the diagram $[\ref{fig:diagramsrg}c]$.

We emphasise that the phenomenology of the novel disorder-driven transition predicted here is similar to the phenomenology of the
non-Anderson disorder-driven transitions~\cite{Syzranov:review} studied previously for systems with isotropic dispersions
$\xi_\bk\propto k^\delta$ in dimensions $\tilde d>2\delta$: renormalised disorder in such systems vanishes for subcritical values of the disorder strength and is finite otherwise.
The RG equations for the flow of the dimensionless disorder strength for such systems are given by
\begin{align}
	\partial_\ell \gamma = (2\delta-\tilde{d})\gamma+\gamma^2+{\cal O}(\gamma^3)
\end{align}
and in one loop are also given by the diagrams shown in Fig.~\ref{fig:diagramsrg}$a^\prime-d^\prime$. We emphasise that for generic symmetries of quenched disorder all of these four diagrams may give contributions of the same order of magnitude to the renormalisation and the higher-loop contributions and in general are non-negligible (see, e.g., Ref.~\onlinecite{Syzranov:TwoLoop}).

The disorder-driven transitions considered in this paper, equivalent to the interaction-driven BEC-vacuum transitions in interacting systems,
are an extension of the previously studied non-Anderson disorder-driven transitions to the case of systems with an anisotropic dispersion $\propto \hsigma_z \xi_\bp+\hsigma_y p_{d+1}$, which is linear along one direction and has a power-law form $\xi_\bp\propto p^\alpha$ along the other $d$ dimensions. The lower-critical dimension for the non-Anderson disorder-driven transitions in such systems is given by $\tilde{d}\equiv d+1=\alpha+1$. The vanishing of the high-order contributions in the RG flow \eqref{RGequation}
is a consequence of the disorder symmetry ($\propto u\hsigma_{z}$) in such systems.


\section{Details of the duality between quantum dot and 1D wires}

\label{Sec:DotWireDualityDetails}

In this section, we provide the details of the duality mapping between the one-sight Hubbard model (quantum dot) described by the Hamiltonian~\eqref{Hdot}
and a disordered 1D wire described by the Hamiltonian~\eqref{Hwires}.
The Hamiltonian of the quantum dot can be rewritten in the equivalent form
\begin{align}
    \cH_{dot}=\xi \hn_\uparrow +\xi \hn_\downarrow -g\hn_\uparrow\hn_\downarrow=\left(\xi+\frac{g}{2}\right) \hn_\uparrow +\left(\xi+\frac{g}{2}\right) \hn_\downarrow -\frac{g}{2}\left(\hn_\uparrow+\hn_\downarrow\right)^{2},
    \label{HamiltonianDotAppendix}
\end{align}
where we have used that $\hn_{\uparrow,\downarrow}^{2}\equiv\hn_{\uparrow,\downarrow}$. 
Observables in a system described by the Hamiltonian \eqref{HamiltonianDotAppendix} can be represented in the form of a path integral over Grassmann variables
\begin{align}
	\langle\ldots\rangle=\int \mathcal{D}\bar{\Psi} \mathcal{D}{\Psi}\ldots
	\exp{\left\{-\int_0^\beta\sum_{i=\uparrow,\downarrow}\bar{\Psi}_{i}(\tau) \left[\partial_{\tau}+\xi+\frac{g}{2}\right] \Psi_{i}(\tau) \,d\tau 
		-\frac{1}{2g}\int_0^\beta\left[\sum_{i=\uparrow,\downarrow}\bar{\Psi}_{i}(\tau)\Psi_{i}(\tau)\right]^2\,d\tau\right\}},
\end{align}
where the preexponential $\ldots$ corresponds to the operator of the observable expressed in terms of the Grassmann fields $\bar{\Psi}$ and ${\Psi}$. 
Decoupling the quartic term by a bosonic field $\phi$ gives
\begin{align}
    \langle\ldots\rangle=\int \mathcal{D}\bar{\Psi} \mathcal{D}{\Psi}\ldots\, \exp{\left\{-\int_0^\beta
    	\sum_{i=\uparrow,\downarrow}\bar{\Psi}_{i}(\tau) \left[\partial_{\tau}+\xi+\frac{g}{2}\right] \Psi_{i}(\tau) \,d\tau -\frac{1}{2g}\int_0^\beta \phi^{2}(\tau)\,d\tau\right\}}.
    \label{PathQuantumDot2}
\end{align}
The action describing the observable in Eq.~\eqref{PathQuantumDot2} corresponds to spin-$1/2$ fermions interacting with bosons whose propagator is given by
$\left<\phi(\tau)\phi(\tau^\prime)\right>=g\delta\left(\tau-\tau^\prime\right)$.


Applying the duality transformation developed in this paper, this model may be mapped to a 1D model with quenched disorder, with the Matsubara time $\tau$ mapped to the coordinate $x$ of the $1D$ model and with the bosonic field $\phi(\tau)$ mapped to a random potential $u(x)$. The Hamiltonian of this 1D model is given by
\begin{align}
	\cH_{wire}=\sum_{i=\uparrow,\downarrow}\hPsi_i(x)\left[\xi\hsigma_z-i\hsigma_y\partial_x+u(x)\hsigma_z\right]\hPsi_i(x).
	\label{HamiltonianWiresAppendix}
\end{align}

\begin{figure}[th!]
	\centering
	\includegraphics[width=\linewidth]{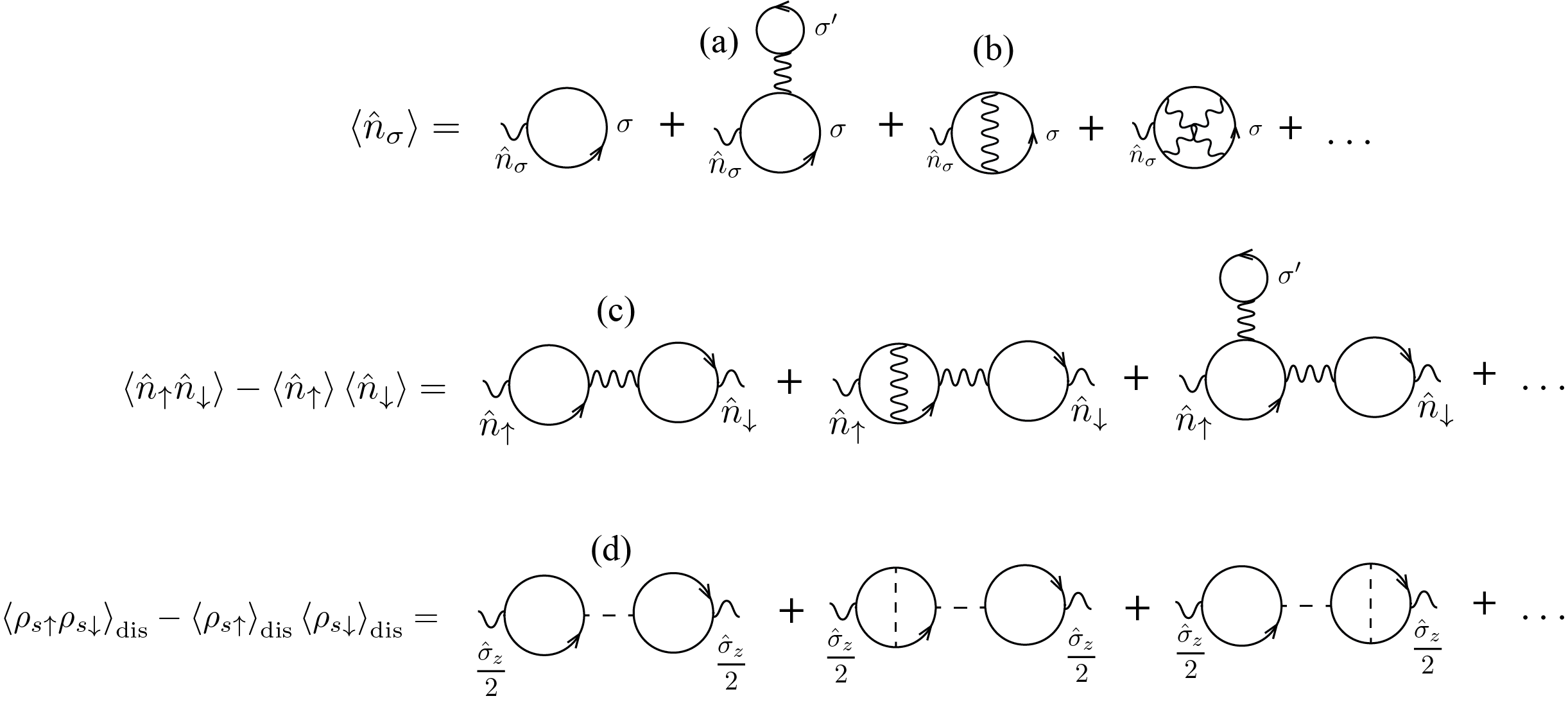}
	\caption{\label{fig:diagramcorrelator} Diagrams that contribute to the correlators $K$ and $K_{\dis}$ in the one-site Hubbard model and a disordered wire described by the Hamiltonians \eqref{HamiltonianDotAppendix} and \eqref{HamiltonianWiresAppendix}.
	}
\end{figure}

We emphasise that, strictly speaking, the quantum dot described by the Hamiltonian~\eqref{HamiltonianDotAppendix} does not satisfy the assumptions
about the negligibility of screening and Hartree-type contributions, which correspond to diagrams with additional fermionic loops and are neglected in this paper when deriving the equivalence between interacting disorder-free and non-interacting disordered systems. 
For example, diagram (a) in Fig.~\ref{fig:diagramcorrelator} describes the Hartree contribution to the average occupation number $\left<\hn_\sigma\right>$ for the electron state with spin $\sigma$ and is equal to diagram (b), which we take into account when demonstrating the equivalence, and is, therefore, non-negligible.

However, observables in the quantum dot may still be mapped to observables in the disordered wire so long as they are unaffected by the screening and Hartree contributions. 
To illustrate this, we consider the leading contribution to the correlator
\begin{align}
    K=\langle\hat{n}_{\uparrow}\hat{n}_{\downarrow}\rangle-\langle\hat{n}_{\uparrow}\rangle\langle\hat{n}_{\downarrow}\rangle
    \label{KAppendix}
\end{align}
of the occupation numbers with different spins in the quantum dot. In the equilibrium state at temperature $T$, the correlator is given by
\begin{align}
	K &=\frac{\sum_{n_{\uparrow,\downarrow}=0,1} n_\uparrow n_\downarrow
	e^{-\frac{n_\uparrow\xi+n_\downarrow\xi-g n_\uparrow n_\downarrow}{T}}}
	{\sum_{n_{\uparrow,\downarrow}=0,1}
		e^{-\frac{n_\uparrow\xi+n_\downarrow\xi-g n_\uparrow n_\downarrow}{T}}}
	-\left(\frac{\sum_{n_{\uparrow,\downarrow}=0,1} n_\uparrow
		e^{-\frac{n_\uparrow\xi+n_\downarrow\xi-g n_\uparrow n_\downarrow}{T}}}
	{\sum_{n_{\uparrow,\downarrow}=0,1}
		e^{-\frac{n_\uparrow\xi+n_\downarrow\xi-g n_\uparrow n_\downarrow}{T}}}
	\right)^2
	\\
	& \approx
	\frac{g}{T}\left[\frac{e^{\xi/T}}{(1+e^{\xi/T})^{2}}\right]^{2},
\end{align}
where we kept only the leading in $g$ contribution.
The correlator~\eqref{KAppendix} can also be found diagrammatically, as shown in Fig.~\ref{fig:diagramcorrelator}. 
The leading in the coupling $g$ contribution is given by diagram (c): 
\begin{align}
    K\approx\quad\includegraphics[width=0.25\linewidth, valign=c]{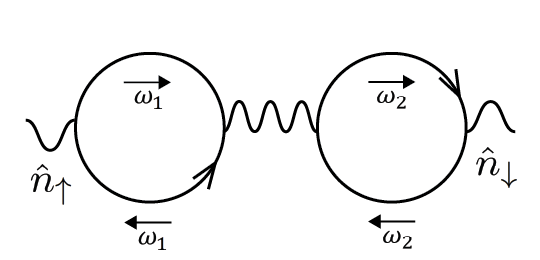}\,\,=g T^{3}\sum_{\omega_1,\omega_2}\frac{1}{(i\omega_1-\xi)^2}\frac{1}{(i\omega_2-\xi)^2}=\frac{g}{T}\left[\frac{e^{\xi/T}}{(1+e^{\xi/T})^{2}}\right]^{2}.
    \label{AppendixK}
\end{align}

Because this contribution does not contain fermionic loops mimicking the screening of the interactions or Hartree contributions, it allows for a mapping to a similar correlator
\begin{align}
    K_{\dis}=\langle{\rho}_{s\uparrow}{\rho}_{s\downarrow}\rangle_{\dis}-\langle{\rho}_{s\uparrow}\rangle_{\dis}\langle{\rho}_{s\downarrow}\rangle_{\dis}.
\end{align}
in a disordered wire described by the Hamiltonian~\eqref{HamiltonianWiresAppendix}.
The diagrams for the correlator in the disordered system are shown in Fig.~\ref{fig:diagramcorrelator}, where the leading-order contribution is given by diagram (d):
\begin{align}
    K_{\dis}&\approx\quad\includegraphics[width=0.26\linewidth, valign=c]{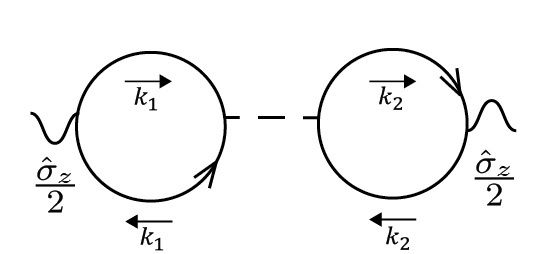}\nonumber\\
    &=\frac{g}{\ell_{d+1}^{3}}\sum_{k_1,k_2}\mathrm{Tr}\left[\hsigma_z\frac{1}{-k_1\hsigma_y-\xi\hsigma_z} \frac{\hsigma_z}{2} \frac{1}{-k_1\hsigma_y-\xi\hsigma_z}\right]\,\mathrm{Tr}\left[\hsigma_z\frac{1}{-k_2\hsigma_y-\xi\hsigma_z} \frac{\hsigma_z}{2} \frac{1}{-k_2\hsigma_y-\xi\hsigma_z}\right]\nonumber\\
    &=\frac{g}{\ell_{d+1}^{3}}\sum_{k_1,k_2}\frac{-k_{1}^{2}+\xi^{2}}{\left(k_{1}^{2}+\xi^{2}\right)^{2}}\,\frac{-k_{2}^{2}+\xi^{2}}{\left(k_{2}^{2}+\xi^{2}\right)^{2}}=g\,\ell_{d+1}\left[\frac{e^{\xi\ell_{d+1}}}{(1+e^{\xi\ell_{d+1}})^{2}}\right]^{2}.
    \label{AppendixKdis}
\end{align}
Because the quantum dot described by the Hamiltonian~\eqref{HamiltonianDotAppendix} is fermionic, the dual disordered wire described by the Hamiltonian~\eqref{HamiltonianWiresAppendix}
has antiperiodic boundary conditions.
At $\ell_{d+1}=1/T$, Eqs.~\eqref{AppendixK} and \eqref{AppendixKdis} for observables in, respectively, the quantum dot and the disordered wire are equivalent, which illustrates again the interactions-disorder duality shown in this paper.


\twocolumngrid


\end{document}